\documentclass[twocolumn,prl,aps,showkeys,showpacs]{revtex4}

\usepackage{graphics}
\usepackage{mathptmx}      
\usepackage{draftcopy}
\usepackage{amsmath}
\usepackage{amssymb}
\usepackage{bm}
\usepackage[usenames]{color}
\usepackage[dvips,colorlinks=true,breaklinks=true]{hyperref}
\hypersetup{%
  pdftitle={Observation of magnetocoriolis waves in a liquid metal
  Taylor-Couette experiment},
  pdfauthor={Mark D. Nornberg},
  pdfsubject={arXiv preprint},
  pdfstartview={FitH},
  linkcolor=MidnightBlue,
  citecolor=BlueViolet,
  urlcolor=Plum,
  anchorcolor=Plum
}
\usepackage{breakurl}

\newcommand{\unit}[1]{\ensuremath{{\hat{\boldsymbol{#1}}}}}

\newcommand{\pder}[3][]{\frac{\partial^{#1} #2}{\partial #3^{#1}}}

\newcommand{\del}{\boldsymbol{\nabla}}
\newcommand{\grad}{\del}
\newcommand{\cross}{\times}

\newcommand{\curl}[1]{\del\cross{ \bf #1 }}

\begin{document}

\title{Observation of magnetocoriolis waves in a liquid metal
  Taylor-Couette experiment}

\author{M.D. Nornberg}
\author{H. Ji}
\author{E. Schartman}
\author{A. Roach}
\author{J. Goodman}

\affiliation{Center for Magnetic Self Organization in Laboratory and
  Astrophysical Plasmas}
\affiliation{Princeton Plasma Physics Laboratory\\
              P.O. Box 451\\
              Princeton, NJ 08543}

\date{\today}

\begin{abstract}
The first observation of fast and slow magnetocoriolis (MC) waves in a
laboratory experiment is reported. Rotating nonaxisymmetric modes
arising from a magnetized turbulent Taylor-Couette flow of liquid
metal are identified as the fast and slow MC~waves by the dependence
of the rotation frequency on the applied field strength. The observed
slow MC~wave is damped but the observation provides a means for
predicting the onset of the Magnetorotational Instability.
\end{abstract}


\maketitle

Hydromagnetic waves are a ubiquitous feature of both geophysical and
solar dynamo models as well as models of astrophysical accretion
disks.  Such waves were first observed in
experiments~\cite{Lundquist.PR.1949} with liquid metals using a
sufficiently strong magnetic field that the Lorentz force could act
like tension on a string and support Alfi{\'e}n
waves~\cite{Alfven.Nature.1942}. When the liquid is in rapid rotation,
these waves are modified by the Coriolis force. The resultant
magnetocoriolis waves~\cite{Acheson.RPP.1973} are a hybrid of
Alfv{\'e}n waves and inertial waves~\cite{Greenspan}. Magnetocoriolis
waves (MC~waves) are used to explain the secular variation of the
Earth's magnetic field over the course of hundreds of
years~\cite{Hide.PTRSL.1966} and the redistribution of angular
momentum in the Sun~\cite{Kumar.APJ.1999}. They are a special case of
the more general Magnetic Archimedes Coriolis (MAC) waves from dynamo
theory~\cite{Roberts.ARFM.1972,Finlay.Dynamos.2008}.

Despite the importance of MC~waves in rotating conducting fluids and
plasmas there is scant experimental evidence of their existence and of
their relationship to various important astrophysical phenomena such
as the dynamo or the magnetorotational instability
(MRI)~\cite{Balbus.AJ.1991A,Sisan.PEPI.2003,Stefani.PRL.2006}. Recent
experiments~\cite{Schmitt.JFM.2008} on a liquid sodium spherical
Couette device and simulations~\cite{Spence.APS.2008} have found
several different types of hydromagnetic waves, but there is ambiguity
about their identification as MC~waves. A similar experiment in
Maryland found inertial waves~\cite{Kelley.GAFD.2007}, but the applied
fields were too weak to observe Lorentz force effects.

In this Letter, we report the first clear identification of the
combined fast and slow MC~waves in a laboratory experiment. Through
measurements of the radial magnetic field in a liquid metal
Taylor-Couette flow, we observe two rotating modes that follow the
rotation speeds expected for the fast and slow MC~wave. We also
demonstrate through a local stability analysis that with the addition
of sufficient flow shear, the slow MC~wave can be destabilized to
produce the MRI. Using the observed frequencies of the waves we infer
from the local dispersion relation that the modes are damped and
obtain a method of determining the threshold for the MRI.

The Princeton MRI experiment is designed to study the stability of a
rotating sheared flow of liquid metal with an applied magnetic field
coaxial with the rotation axis. The apparatus has been described
elsewhere~\cite{Schartman.RSI.2009} and has already demonstrated the
ability to generate high Reynolds number shear flow in water with
angular momentum flux comparable to viscous transport (a null result
in trying to demonstrate subcritical hydrodynamic
instability~\cite{Ji.Nature.2006}).  The volume between the concentric
rotating cylinders is filled with GaInSn, a gallium eutectic
alloy~\cite{Morley.RSI.2008}. The dimensions of the experiment and
properties of the liquid metal are given in Tab.~\ref{tab:parameters}.

\begin{table}
\begin{tabular}{|c|c|c|c|}\hline
Experimental Parameters & Symbol & Units & Value\\ \hline
Height & $h$ & cm & 27.9\\
Inner Cylinder radius & $r_1$ & cm & 7.06\\
Outer Cylinder radius & $r_2$ & cm & 20.30\\
Density & $\rho$& ${\rm g}/{\rm cm}^3$ & 6.36\\
Kinematic viscosity & $\nu$& ${\rm cm}^2/\mbox{s}$ & $2.98 \times 10^{-3}$\\
Magnetic diffusivity & $\eta$ & ${\rm cm}^2/\mbox{s}$ & $2.57 \times
10^3$\\ \hline
\end{tabular}
\caption{Physical parameters of the apparatus~\cite{Schartman.RSI.2009} and
  liquid metal~\cite{Morley.RSI.2008}.}
\label{tab:parameters}
\end{table}

The equations describing the evolution of a rotating shear flow with a
background magnetic field are given by the magnetohydrodynamic (MHD)
equations in a rotating frame:
\begin{eqnarray*}
\pder{\mathbf{v}}{t} + \left( \mathbf{v} \cdot \boldsymbol{\nabla}
\right) \mathbf{v} + 2 \boldsymbol{\Omega} \times \mathbf{v} & = &
- \grad{P} + \frac{1}{\mu_o \rho} \left( \mathbf{B}
\cdot \nabla \right) \mathbf{B} + \nu \nabla^2 \mathbf{v}\\
\pder{\mathbf{B}}{t} & = & \curl{\left(\mathbf{v} \times \mathbf{B}\right)}
+ \eta \nabla^2 \mathbf{B}\\
P & = & \frac{p}{\rho} + \frac{1}{2} \frac{B^2}{\mu_0 \rho} -
\frac{1}{2} \left| \boldsymbol{\Omega} \times \mathbf{r} \right|^2
\end{eqnarray*}
where $\mathbf{v}$ and $\mathbf{B}$ are solenoidal fields representing
the velocity and magnetic field, $\boldsymbol{\Omega}$ is the angular
velocity, and $p$ is the pressure. The generalized pressure $P$
incorporates the magnetic and centrifugal pressure terms.

The background field and angular velocity are given by
$\mathbf{B}_0 = B_0\,\hat{\mathbf{z}}$ and $\boldsymbol{\Omega} =
\Omega\,\hat{\mathbf{z}}$. If we assume harmonic perturbations of
the velocity and magnetic field and linearize the resulting equations,
we obtain the dispersion relation in cylindrical
coordinates~\cite{Ji.MNRAS.2001}:
\begin{equation}
\begin{split}
\left( \bar{\omega} - i\,\gamma_\eta \right)^2 \left[ \left( \bar{\omega}
  - i\,\gamma_\nu \right) \left( \bar{\omega} - i\,\gamma_\eta \right) + 
  \omega_A^2 \right]^2\\ 
+ 2 \zeta \Omega^2 \left( \bar{\omega} -
i\,\gamma_\eta \right)^4 \left( k_z/k \right)^2 - 2(2-\zeta) \Omega^2 
\omega_A^2 \left( \bar{\omega} - i\,\gamma_\eta \right) \left(
k_z/k \right)^2 \\
 - \omega_R \left[ \left( \bar{\omega} -i\,\gamma_\nu \right) \left(
   \bar{\omega} - i\,\gamma_\eta \right) + 
  \omega_A^2 \right] \left[
  \left( \bar{\omega} -i\,\gamma_\eta \right)^2 + \omega_A^2 \right]
 = 0
\end{split}
\label{eq:dispersion}
\end{equation}
where $k_\phi=m/r$ for integer $m$, $\bar{\omega} = \omega -
m\Omega$ is the Doppler-shifted complex frequency, $\gamma_\eta = \eta
k^2$ and $\gamma_\nu = \nu k^2$ are the resistive and viscous damping
rates, $\omega_A = k_z B_0 / \sqrt{\mu_0 \rho}$ is the Alfv{\'e}n
frequency, and $\omega_R = \left( \zeta - 2\right) \Omega k_r k_\phi /
k^2$ is the Rossby wave frequency~\cite{Rossby.JMR.1939}. We
have quantified the flow shear by introducing the vorticity parameter
\begin{equation}
\zeta(r) = \hat{\mathbf{z}} \cdot \frac{\del\times r\Omega(r)
  \unit{\phi}}{\Omega(r)} = \frac{1}{r\Omega}
\frac{\partial(r^2\Omega)}{\partial r}.
\end{equation}
Note that $\zeta$ is only constant if the rotation profile follows a
power-law dependence $\Omega(r) \propto r^{\zeta-2}$.  For uniform
rotation with no shear $\zeta = 2$. The Rayleigh
criterion~\cite{Rayleigh.PRSLA.1917}, which governs axisymmetric
hydrodynamic centrifugal stability of rotating shear flow, is given by
$\zeta \ge 0$ .

\begin{figure}
\includegraphics{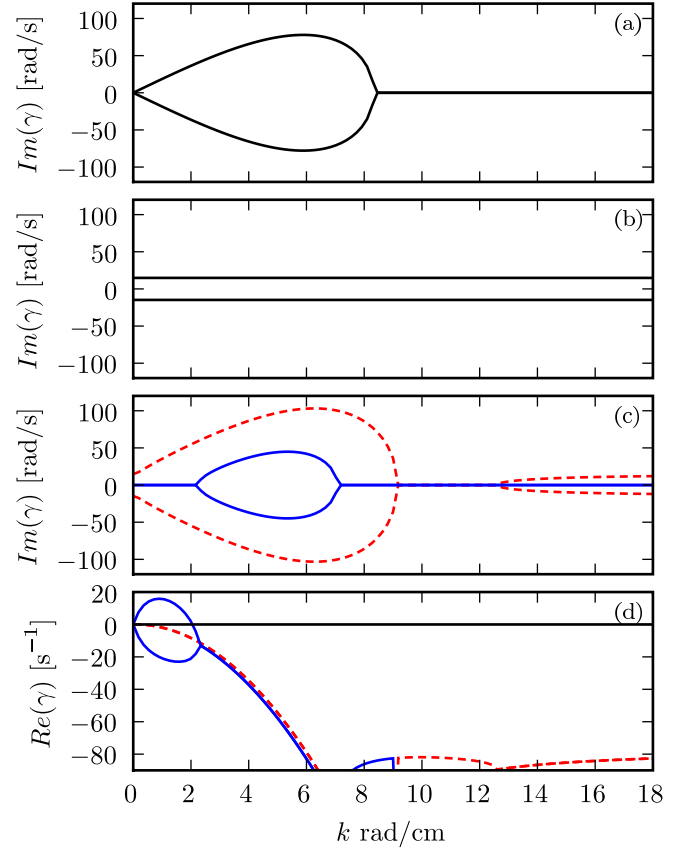}
\caption{The real part of the dispersion relation for (a) damped
  Alfv{\'e}n waves as given in Eq.~\ref{eq:alfven}, (b) inertial waves
  as given in Eq~\ref{eq:inertial}, and (c) the fast (dashed red line)
  and slow MC~waves (solid blue line). (d) The growth rate of the fast
  and slow MC~waves. Due to resistive dissipation, the MC~waves
  collapse to the inertial wave dispersion at high $k$. When there is
  sufficient flow shear, the slow MC~wave becomes unstable at low
  $k$. The parameters for the dispersion relation are $B_0 = 4$~kG,
  $\Omega = 42$~rad/s, and $\zeta = 0.25$ (maximum design
  parameters of the apparatus) and the fluid properties are
  given in Tab.~\ref{tab:parameters}.}
\label{fig:dispersion}
\end{figure}

We can gain insight into the basic waves for this dispersion relation
by examining limiting cases. In the absence of rotation,
the dispersion relation reduces to
\begin{equation}
\left( \omega - i \gamma_\nu \right) \left( \omega -i \gamma_\eta \right) 
- \omega_A^2 =0
\label{eq:alfven}
\end{equation}
which describes the damped shear Alfv{\'e}n wave. These waves have a
transverse polarization due to the incompressibility of the fluid (in
a compressible fluid there is also a longitudinal magnetosonic
wave). The complex frequency is $\omega = i (\gamma_\nu+\gamma_\eta)/2
\pm \sqrt{\omega_A^2 - \left(\gamma_\nu - \gamma_\eta\right)^2}$ which
shows that the Alfv{\'e}n wave is viscously and resistively
damped and has a real frequency of $\pm \omega_A$ in the absence of
dissipation.  Variations in the flow along the direction of the
magnetic field tend to be eliminated in a highly resistive fluid (such
as a liquid metal) due to this damping.

Assuming rotation without shear and without an applied magnetic field,
the dispersion relation reduces to
\begin{equation}
(\omega - i\,\gamma_\nu )^2 + (2\Omega k_z/k)^2 = 0
\label{eq:inertial}
\end{equation}
which describes inertial waves. Inertial waves also have a transverse
polarization but are peculiar in that the restoring force, provided by
the Coriolis effect, acts orthogonally to the displacement. The
resulting motion of a displaced fluid element is circular
precession. The complex frequency is $\omega = i\gamma_\nu \pm
2\Omega k_z/k$ so the wave is viscously damped with a real
frequency between $\pm 2\Omega$. Note that there is no dependence on
the wavelength. Akin to Alfv{\'e}n waves, inertial waves homogenize
the flow along the axis of rotation due to viscous damping, consistent
with the Taylor-Proudman Theorem~\cite{Greenspan}.

\begin{figure}
\includegraphics{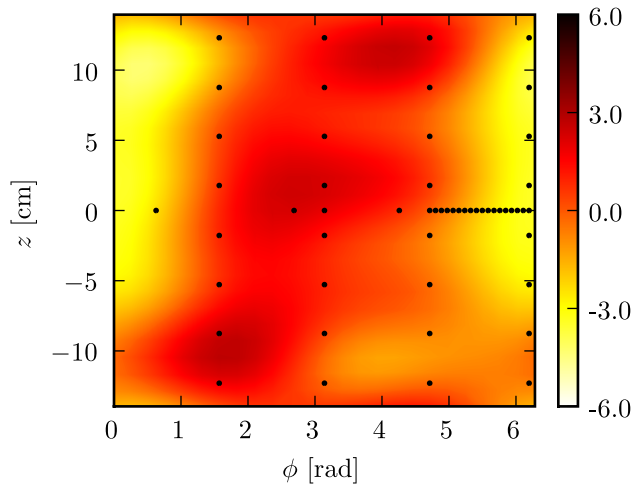}
\caption{A contour plot of the radial magnetic field near the surface
  of the outer cylinder constructed from a least squares fit of data
  from the Mirnov coil array. The locations of the coils are depicted
  by dots and the contour levels are given in Gauss. The applied field
  was 4.3~kG.}
\label{fig:mode}
\end{figure}

\begin{figure*}
\includegraphics{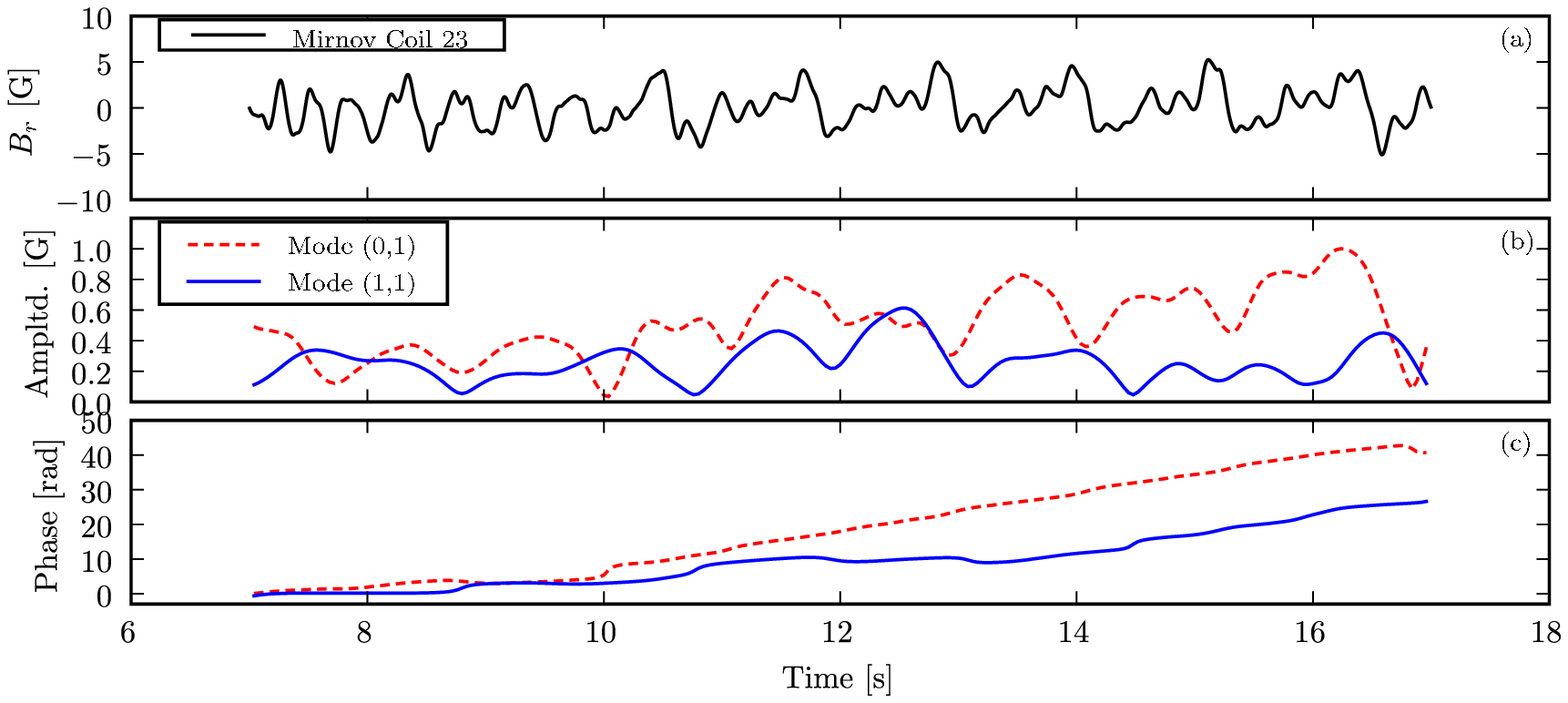}
\caption{(a) Magnetic field measured by a single Mirnov coil. (b) The
  time series of (0,1) and (1,1) mode amplitudes from a 2D Fourier
  decomposition of the radial magnetic field where the notation
  ($n,m$) describes the vertical mode number $n$ and the azimuthal
  mode number $m$. (c) The time series of mode phases.}
\label{fig:summary}
\end{figure*}

Together these waves make up the hybrid magnetocoriolis
wave~\cite{Acheson.RPP.1973}.  Since there are two restoring forces
acting on displaced fluid elements, there are two possible
situations. The Lorentz and Coriolis forces may act together,
stiffening the system and producing the higher frequency fast wave, or
the two forces may oppose one another to produce the lower frequency
slow wave. The requirement for observing strong rotational effects on
the Alfv{\'e}n wave is $r \Omega \sqrt{\mu_0 \rho}/ B_0 \sim 1$. The Alfv{\'e}n
frequency experiences a splitting due to the breaking of the
degeneracy of the two roots by the presence of rotation. When the flow
has sufficient shear, the slow MC~wave becomes stationary and unstable
at low $k$ as seen in Fig.~\ref{fig:dispersion}. This instability is
the MRI which in addition to causing turbulent transport of angular
momentum in accretion disks may also be related to geomagnetic
jerks~\cite{Petitdemange.GRL.2008}.

The results reported here were obtained when an axial magnetic field
was applied to turbulent rotating shear flow. The outer cylinder was
kept stationary while the inner cylinder rotated. The rings at the end
caps were coupled to either cylinder in what is referred to as the
``split'' configuration~\cite{Schartman.RSI.2009}. The flow was
hydrodynamically unstable since $\zeta \leq 0$. The inner cylinder and
inner ring were set in rotation at 6.7~Hz and an axial magnetic field
between 1.7 to 4.3~kG was applied to the turbulent flow. The induced
radial magnetic field fluctuations were measured by an array of Mirnov
coils\cite{Hutchinson.2002} positioned just outside the outer
cylinder. A nonaxisymmetric mode was apparent from the coil array as
seen in the snapshot of the field shown in Fig.~\ref{fig:mode}. The
image is created by fitting the signals from the array to a spatial
Fourier mode model using standard least squares fitting. From the
model we obtain the Fourier amplitude and phase as shown in
Fig.~\ref{fig:summary}. The variation of the mode amplitudes with
magnetic field strength is not linear as would be expected for
advection by differential rotation (the $\Omega$~effect).

\begin{figure}
\includegraphics{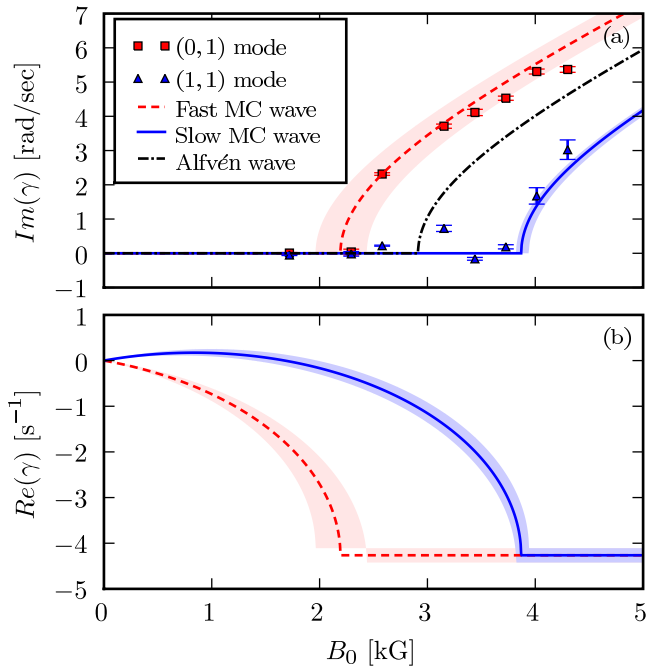}
\caption{The real frequency (a) and growth rate (b) determined by
  fitting the phase speed of the two nonaxisymmetric modes observed
  for a range of applied magnetic field strengths to
  Eq.~\ref{eq:dispersion}. Error bars reflect the uncertainty in the
  linear fit to the phase as a function of time. The dashed red and
  solid blue lines show the fit of the fast and slow MC~waves from
  Eq.~\ref{eq:dispersion} with values of $\zeta=0$, $k=0.246 \pm
  0.001\ {\rm cm}^{-1}$, $\theta = 1.336 \pm 0.007\ {\rm rad}$ where
  $k_z = k \cos\theta$, and $\Omega = 5.4 \pm 0.9\ {\rm rad/s}$. The
  shaded areas express the uncertainty in the fit.}
\label{fig:phase_match}
\end{figure}

The mode rotation rates are measured by calculating the linear slope
of the Fourier phase with time for each field strength. The results
are shown in Fig.~\ref{fig:phase_match}. The modes clearly rotate at
different speeds and their rotation rates increase with magnetic field
strength. By comparing the rotation rates with those of higher
azimuthal harmonics obtained from the high density midplane coil
array (seen as the series of dots at $z=0$ in Fig.~\ref{fig:mode}) we
find that the harmonics are not phase locked. Hence, the signal is not
likely due to a passing vortex as was observed in
~\citet{Sisan.PEPI.2003}. They are also not the result of the MRI
since the least-damped mode should be axisymmetric.

A least squares fit of the observed mode rotation rates to the real
frequencies of the fast and slow MC~waves gives an estimate of the
local wave vector components and the fluid rotation rate for a given
shear. The wavenumber components are fit since the observed magnetic
field is much smoother than the velocity field due to the low $Pm$ and
cannot reveal detailed structures like the boundary layers. The fit is
insensitive to the vorticity parameter $\zeta$ and so we assume it to
be zero for marginal stability consistent with hydrodynamic
observations~\cite{Lewis.PR.1999}. The fit parameters result in a mode
with $k_z = \pi/2h$ and $k_r = \pi/(r_2-r_1)$ which corresponds to the
smallest (and therefore least damped) radial wavenumber that can fit
in the radial gap and a quarter-wavelength vertically. The growth rate
determined from the fit is shown in Fig.~\ref{fig:phase_match}b. Aside
from an expected Doppler shift for the nonaxisymmetric modes, the
MC~wave model provides an excellent fit to the observations. From the
growth rate, we find that the slow wave has a small positive growth
rate for 1~kG but is otherwise damped. The model suggests that the MRI
growth rate is too small at the rotation rates achieved to observe it
with our diagnostics. By adjusting the model parameters we can predict
the necessary rotation rate and magnetic field strength to observe the
MRI based on the empirical observations of damped waves.

Nonaxisymmetric waves have been observed in the PROMISE magnetized
Taylor-Couette experiment~\cite{Stefani.JP.2007} which has also
observed the helical
MRI~\cite{Stefani.PRL.2006,Rudiger.APJ.2006}. \citet{Sisan.PRL.2004}
also observed rotating nonaxisymmetric spherical harmonic patterns in
a turbulent liquid sodium spherical Couette flow which they attribute
to the MRI, but may be due to a shear instability of the secondary
meridional flow~\cite{Hollerbach.PRSA.2009}. Runout of the inner
cylinder could lead to a spinover mode due to the elliptical
instability, but it is suppressed by the magnetic
field~\cite{Herreman.PF.2009}. It is still unclear why these
nonaxisymmetric waves are favored over the axisymmetric ones, and its
resolution may require 3D simulations as well as internal flow and
magnetic field measurements.

One conjecture regarding the source of these damped waves is
that the turbulent flow provides perturbations of a broad range of
wavelengths, but that the geometry of the vessel dictates which modes
are realized~\cite{Wood.JFM.1965}. Such is observed when a precessing
top cap is used to drive inertial waves in a cylinder filled with
water~\cite{McEwan.JFM.1970}. The observed waves in this case are
cavity resonances driven by unstable flow at the boundary
discontinuities and not an instability of the bulk flow. It is also
possible that these damped waves are driven through nonlinear wave
coupling~\cite{Terry.PRL.2004}. If such is the case, then these damped
waves may be a saturation mechanism for the MRI. Although we have not
observed the MRI in the experiment, nonlinear simulations of single
mode MRI found that saturation was achieved by amplifying the vertical
field through an $\alpha$~effect~\cite{Ebrahimi.AJ.2009}. It is due
to the ambiguity of the source of these waves that we are continuing
to pursue an observation of magnetically induced instability in a
hydrodynamically quiescent flow.

In summary, we have observed rotating modes in a turbulent
Taylor-Couette flow of liquid metal that we identify as the fast and
slow MC~wave. We have identified a relationship between the slow
MC~wave and the MRI and have proposed a method of determining the
threshold for instability through observation of driven
MC~waves. MC~waves will be important in identifying the MRI in further
experiments and may also play a role in saturation of MHD turbulence
in rotating fluids.

We thank Chris Finlay for helpful discussions and pointing out
valuable references and Hans Rinderknecht for his contributions on
dispersion relation calculations as a senior project of Princeton
University. This work was funded under NASA Grants No.
ATP03-0084-0106 and No. APRA04-0000-0152, DOE Contract No.
DE-AC02-09CH11466, and NSF Grant No. AST0607472.


\end{document}